# A Review of Liquid Phase Epitaxial Grown Gallium Arsenide

D. Alexiev, D. A. Prokopovich, S. Thomson, L. Mo, A. B. Rosenfeld and M. Reinhard

## *PART 1*

## *ABSTRACT*

Liquid phase epitaxy of gallium arsenide (LPE GaAs) has been investigated intensively from the late 1960's to the present and has now a special place in the manufacture of wide band, compound semiconductor radiation detectors. Although this particular process appears to have gained prominence in the last three decades, it is interesting to note that its origins reach back to 1836 when Frankenheim made his first observations.

A brief review is presented from a semiconductor applications point of view on how this subject developed. This is followed by a report on LPE GaAs growth at the Australian Nuclear Science and Technology Organisation (ANSTO).

## *HISTORICAL INTRODUCTION*

Liquid phase epitaxy (LPE) is the solidification from a liquid phase of a crystalline layer onto a parent substrate such that the crystallinity of the substrate is maintained in the grown layer. Such crystal growth has received continuing attention since first observed in natural formations. These observations lead to experimental studies and it appears that Frankenhein (1836) was the first in a long line of crystal growers to grow LPE layers. Frankenhein found that sodium nitrate grew from solution in an oriented direction on the surface of freshly cleaved calcite crystals. Baker (1906) extended this early work with a series of systematic experiments in which a number of related structures were grown upon each other. Baker's method consisted of placing a drop of saturated solution of alkalides onto a cleaved surface and observing the nucleation of crystalline structure through a microscope. Royer (1928) continued Baker's work, and work with the aid of the newly discovered X-ray diffraction analysis of structures, greatly increased the scope of studies of the epitaxial layers. Royer developed rules for epitaxy, of which the most important is that oriented growth occurs only when it involves the parallelism of two lattice planes which have lattice networks of identical or quasi-identical form and closely similar spacings. The term 'percentage misfit' evolved, referring to the differences between the lattice network spacings or "lattice parameters'. Royer found experimentally that lattice-parameter misfit should be no more than 15 as demonstrated by the growth of alkali halides upon other alkali halides and on mica. Electron diffraction studies by Finch and Quarrell (1933) added a further insight into epitaxial growth and misfit. They showed that growth can occur with an initial oriented film which has a modified crystalline structure. The bulk structure of the epitaxy is then constrained so that the lattice plane parallel to the substrate remains identical in size. J.H. van der Merwe (1949) continued with this approach and developed a theoretical approach to epitaxy and the formation of the 'misfit' layer. His theory also predicted a limit, in magnitude, to the misfit of the lattice network beyond which epitaxial growth cannot proceed, similar, as had been found by Royer. Later, in a detailed review D.W. Pashley (1956) concludes that a small misfit is not an essential criterion for epitaxial crystal growth to occur. He finds Royer's results very convincing in that the misfit value is significant under certain conditions. However, the theoretical derivation by van der Merwe - the concept of pseudomorphic monolayers - is not correct for many cases of epitaxy. Pashley stresses that chemically grown deposits require special attention since the substrate undergoes changes during the growth of a surface layer; requiring both experimental and theoretical studies of the nucleation problem. It is interesting to note that the special conditions of substrate melt-back and super cooling during the initial epitaxial growth stages have not yet been introduced by experimentalists.

The modest but continuing interest in epitaxy, and in particular liquid phase epitaxy, changed abruptly with the development of the semiconductor industry in the early 1960s. Semiconductor technology at that time was based entirely on germanium with silicon becoming dominant later. However, it was found that Ge and Si had certain limitations for particular device construction. Their band gaps are indirect and are fixed at $Eg(Ge) \approx 0.68$ eV and $Eg(Si) \approx 1.1$ eV. Hence, they are not useful as light emitters, transferred electron devices (Gunn oscillators) or efficient solar energy converters (Holonyak et al (1978)). When constructing room temperature operating radiation detectors, high band gaps are required such as $Eg(GaAs) \approx 1.41$ eV to reduce thermally generated leakage currents and high purity with very low carrier concentration is needed to create large depletion volumes in these devices.

In general, it is these special properties in the III-V semi-conductor materials, which led to concentrated research activities from 1966 to the mid-seventies into liquid phase epitaxial layer growth.

## MODERN APPROACH TO LPE GROWTH OF GALLIUM ARSENIDE

A successful and simple method (Figure 1) for growing LPE GaAs crystals was introduced by H. Nelson (1963). This involved heating a GaAs seed (substrate) next to a solution of tin-GaAs mixture placed at the lower end of a graphite crucible. The graphite crucible is then heated to about 640°C. When the furnace reaches the selected temperature (Figure 2), the power is turned off and the furnace tipped so that the molten tin covers the exposed surface of the GaAs wafer. When cooled to about 400°C, the furnace is tipped back to its original position. Immediately afterwards the graphite crucible is removed and any remaining tin is wiped off the epitaxial layer.

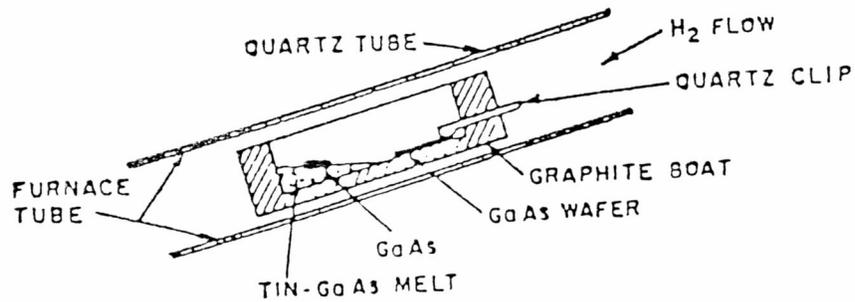

Figure 1 – Apparatus for LPE growth of GaAs from a tin solution (after Nelson, 1963)

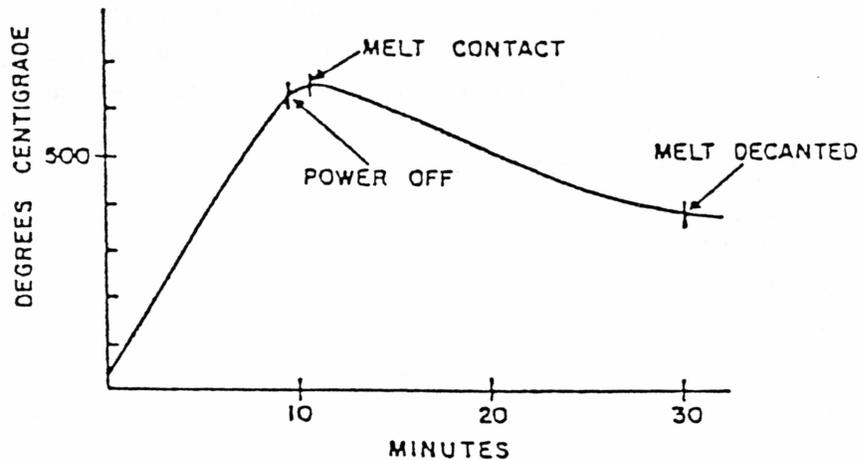

Figure 2 – Heating schedule for epitaxial deposition of GaAs from a tin solution (after Nelson, 1963)

Nelson reported epitaxiel layers that were typically 60 to 80 μm thick. Surfaces were rough due to rapid growth in the low temperature regime near the completion of the process and large (in the order of several μm) droplets of Ga were also noted at the interface. A.R. Goodwin et al (1968) recognised this problem of rapid growth. Their solution was to introduce a temperature gradient, approximately 10°C cm$^{-1}$ so that the seed was always colder than the melt by a fixed amount. As before, the boat was tipped at 850°C and the furnace temperature lowered at a rate of 10°C/min to 600°C. Thus layers of 150 to 200 μm thickness were grown over 3 hours. The surfaces of the epilayer were good and occasionally mirror bright. Such experimental work led Goodwin et al to reconsider the travelling solvent technique (TST) reported earlier by Miavsky and Weinstein (1963). The Ga solvent is saturated with As by adding GaAs crystals, some are dissolved while others remain floating on the surface throughout the growth cycle.

Difficulties were encountered when the melt did not melt the surface of the seed, a problem apparently overcome by preliminary baking of the seed in vacuum at 800°C.

The role of constitutional supercooling in the solution growth of GaAs was first reported by Tiller (1968). He pointed out that if the temperature gradient at the interface is insufficient, then constitutional supercooling will occur, rendering the interface unstable and resulting in both an uneven surface quality and gallium inclusions. Taller thus derived minimum required values for the temperature gradient in a steady state, diffusion limited growth process. This work was extended by Minden (1969) who derived detailed diffusion equations for the minimum allowable temperatures to avoid constitutional supercooling. In the same year Hicks and Manley (1969) used Nelson's method to produce LPE GaAs with exceptionally low carrier concentrations in the order of $10^{12}$ cm$^{-3}$. A maximum mobility of $2.5 \times 10^5$ cm$^2$/V-sec at 51K was reported. Two important steps were introduced, firstly the melt was baked at 850°C for 14 hr under a flow of H and secondly, the purity of the feed material was maximised to a net carrier concentration of $10^{15}$ cm$^{-3}$ and a copper content not greater than 0.1 part per million, the solvent Ga was 99.9999% pure.

Following tipping, the furnace was linearly cooled at 25°C h$^{-1}$. The substrate used was a semi-insulating Cr-doped single crystal with a <100> orientation. Hicks and Manley noted that some of the LPE layers were highly non-uniform in mobility, suspecting inhomogenity (off-stoichiometry) in the LPE rather than a variation in impurity concentration.

It is interesting to speculate that the non-uniformity of mobility could have been due to substrate melt-back into the Ga solution during tipping, despite the melt being saturated at the growth starting temperature. Freed Cr could then have formed semi-insulating regions in the epitaxial growth front.

In a following publication Hicks and Greene stressed the importance of the quality of the hydrogen atmosphere used during the LPE growth; they produced a theoretical study, with experimental support, of the evolution of free Si (now known to be a shallow acceptor level in LPE GaAs, concentration in the Ga solvent. Hicks and Greene found a direct proportionality between free Si concentration in the Ga solvent and the net hole concentration in the epilayer. They showed that silica boats are reduced by the hydrogen atmosphere, introducing free Si into the Ga solvent as a significant contaminant. When a small controlled quantity of $O_2$ is introduced into the gas stream, $H_2O$ is produced in the hot furnace so that the free Si contamination in the melt is limited to an acceptable level. The equilibrium concentration of Si in the Ga melt (solvent) is expressed simply as

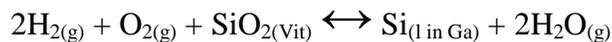

$$2H_{2(g)} + O_{2(g)} + SiO_{2(Vit)} \longleftrightarrow Si_{(l\ in\ Ga)} + 2H_2O_{(g)}$$

(1)

This crucial reaction does not involve As. Ga appears only indirectly as a solvent for the silicon. Hicks and Greene noted that such control of silicon contamination is not only relevant to the growth of LPE of GaAs but is important for the growth of any gallium compound by the same technique. By applying these principals Hicks and Greene produced epitaxial layers of remarkably low carrier concentrations in the $10^{13}$ cm$^{-3}$ range and high electron mobility $\mu^e = 100$ k cm$^2$ v$^{-1}$ s$^{-1}$. Eberhardt et al constructed a high resolution X-ray detector from samples of this material. The device was a simple surface-barrier radiation detector in which a measure of the carrier concentration confirmed a value of $2 \times 10^{-13}$ cm$^{-3}$. However, they also found a typical epitaxial-substrate interface (ESI) layer ranging in thickness up to 2 nm. These ESI layers represent an anomalous discontinuity in the carrier concentration acting as a semi-insulating layer. This ultimately degrades the performance of the device if the substrate is to be used as the ohmic contact. When used as an X-ray spectrometer to resolve $^{241}$Am, a resolution of 640 eV FWHM was obtained for the 59.54 keV γ line. The detector was cooled to 122 K to reduce reverse leakage current ($I_R$) and thereby optise resolution. A $^{57}$Co spectrum at room temperature produced 2.6 keV FWHM for the 122 keV line. It is interesting to note that the resolution obtained by Eberhardt el al (1971) has not since been bettered in LPE GaAs.

A further insight into the behaviour of Si in GaAs has been gained by M.E. Weiner (1971). A model was proposed based on the formation of silicon-oxygen pairs to explain a variety of anomalous behaviour of LPE GaAs when grown with Si, SiO$_2$ and O$_2$. Weiner suggested that Si atoms on Ga sites paired with interstitial oxygen atoms. They form a complex which behaves as an acceptor with energies of 0.1 and 0.4 Ev below the conduction band. The complex is assumed to dissociate upon annealing below 850°C by the reaction:

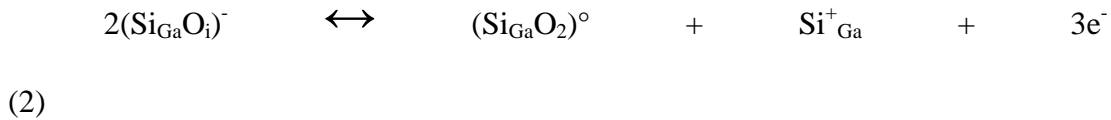

$$2(Si_{Ga}O_i)^- \leftrightarrow (Si_{Ga}O_2)° + Si^+_{Ga} + 3e^-$$

(2)

At higher temperatures this reaction may be reversed, thereby explaining changes from n to p type conductivity as a function of Si concentration in GaAs grown from Ga solution in a silicon boat. This model contradicts that of Hicks and Greene, which simply concludes that addition of oxygen, slows the reduction process of the silicon boat by the H atmosphere and ultimately lowering the acceptor density in the epilayer.

In a further study of the problem of Si contamination of GaAs, Weiner (1972) proposed three specific cases, of which the first two, shown in Figure 3, are of particular relevance to LPE growth:

Case A: The contamination from an inert crucible or boat in a flow of H$_2$ : here the Si incorporation into the Ga liquid becomes significant only at larger temperatures (> 800°C) and in a very dry H$_2$ flow over long periods. Weiner's calculations were based on local thermodynamic equilibrium when the initial pressure of water ($P_{H_2O°}$) in the system, including any water from reaction between H$_2$ and O$_2$, is increased. He found that the rate of contamination depends only slightly on $P_{H_2O°}$ but 2 when the $P_{H_2O°}$ is increased by the introduction of O$_2$ then the rate of Si contamination drops rapidly; this is because a significant fraction of total

H$_2$O content is generated by the H$_2$ reduction of the SiO$_2$ furnace tube in a very dry system.

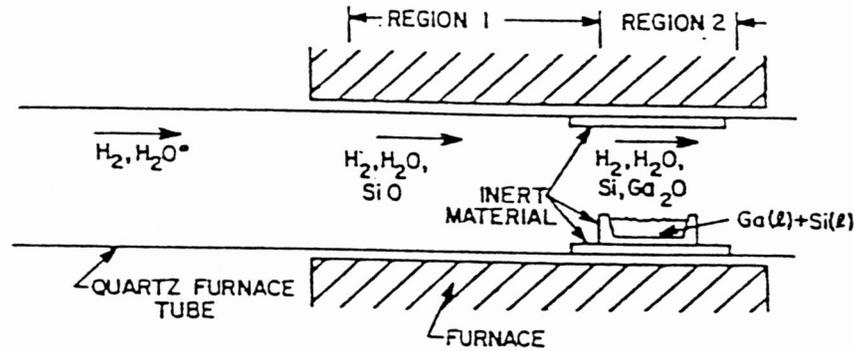

Figure 3 – Case A

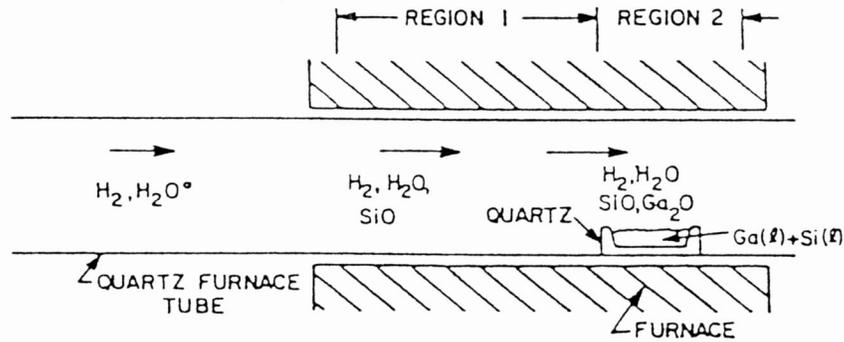

Figure 3 – Case B

Case B: Si contamination when the Ga is located in a quartz crucible (such as used by Hicks and Green and by the author). Most of the Si contamination in the Ga liquid is provided by the Ga reduction of its quartz crucible. The Si contamination rises very rapidly under dry H$_2$ conditions, such that after 1 hr at 1000°C 10 ppm of Si can be expected in the Ga. However, if more than 10$^{-4}$ atm pressure of water vapour is added to the system, the Si steady-state value is significantly reduced. An interesting observation made by Weiner is that there is a significant decrease in the Si steady-state concentration if the H$_2$ is replaced with an inert gas. P.B. Greene (1973) considerably simplified Weiner's kinetic and thermodynamic calculations. Essentially Greene used the same approach as earlier (Hicks and Greene) but extended the calculation to the rate at which Si concentration changes in the melt as well as the steady state equilibrium concentration. He showed that, in particular, it is H$_2$O rather than Ga$_2$O that is the predominant species involved in O$_2$ removal from the crucible vicinity (<1000°C). This means that the rate at which reduction occurs and at which Si enters the Ga liquid, is determined by the rate at which H$_2$O concentration up stream [H$_2$O input] minus the H$_2$O downstream and out of the crucible can be removed by the H flow. Green maintains that his earlier, extremely high purity LPE-GaAs (N$_D \approx$ 10$^{13}$ cm$^{-3}$), growth results was only possible when water vapour was added to the H stream and when the growth was commenced at a particular temperature. An appreciation of this condition can be seen from the simple differential equation, (adapted from Greene) expressing the rate at which Si enters the Ga liquid.

Rate at which Si enters the melt:

$$\frac{dN_{Si}}{dt} = \frac{([H_2O_{OUT}] - [H_2O_{IN}])F}{2RT_m}$$

$$\frac{dN_{Si}}{dt} = N_{Ga}\frac{d[Si]}{dt}$$

(Since in the temperature range of interest the melt consists mostly of Ga), and on the assumption that $H_2O$ downstream is the equilibrium value for the solution in the crucible:

$$\frac{d[Si]}{dt} = \frac{\{K_R^*/[Si]^{1/2} - [H_2O_{IN}]\}}{2RT_m N_{Ga}}$$

When the silicon concentration is very small, the term [$H_2O$ in ] is negligible compared with the term $K_R^*/[Si]$) so that the above expression reduces to

$$\frac{d[Si]}{dt} = \frac{K_R^*/[Si]^{1/2} F}{2RT_m N_{Ga}} \qquad \text{(Where F = gas flow and K = [Si] [H}_2\text{O]}^2\text{)}$$

With the additional condition that [Si] =0 when t = 0, the solution of the simplified differential equation is

$$[Si] = K_R^{*1/3}(3F/4RT_m N_{Ga})^{2/3} t^{2/3}$$

The important feature of this solution is that it takes longer to reach equilibrium at higher temperatures since the temperature dependence (arising only from the variation of $K_R^*$ with temperature) of the silicon concentration is much less than the temperature dependence of the final (equilibrium) value, which depends on $K_R^*$ to the power of one.

To illustrate this condition, Greene evaluated the differential equation for temperatures of 800°C (Figure 4a) and 1000°C (Figure 4b) at various water concentrations with a flow of 1 litre min$^{-1}$ of $H_2$ and 25g of gallium melt.

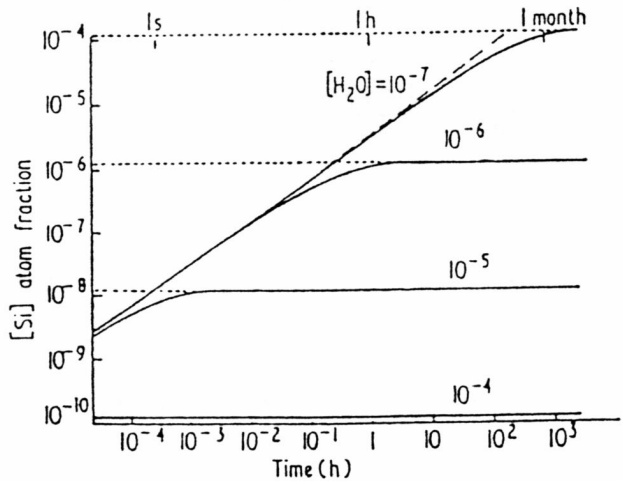

Figure 4a – Silicon contamination of liquid gallium in a silver boat at 800°C

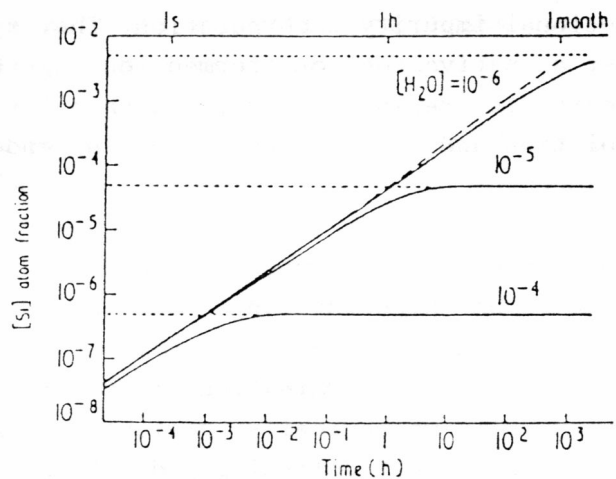

Figure 4b – Silicon contamination of liquid gallium in a silicon boat at 1000°C (Green, 1973)

The graphs show that high levels of silicon contamination arising from the use of high temperatures and low water concentrations in the gas stream require extremely long times to approach equilibrium. When conditions are chosen to limit the silicon contamination of the melt to less than 1 part in $10^7$, the equilibrium concentrations are attained rapidly.

Epilayers between 0.5 to 1μm thickness are required for construction of devices such as bipolar GaAs transistors, FETs, LEDs and Gunn diodes. The Nelson tilt-tube furnace is not suitable, since the surface morphology, uniformity and thickness are not reproducible. For this reason Vilms and Garrett (1971) introduced a graphite crucible which minimised the retention of Ga droplets, facilitating thickness control by allowing continuous agitation of the Ga solution without sliding seals. Vilms and Garrett's main concern was air leaks into the growth apparatus at seals and at the H purifier membrane. They suggested that oxygen is the impurity responsible for acceptor and donor concentrations in the range $10^{15}$ to $10^{16}$ cm$^{-3}$. At higher temperatures, 700 to 850°C then the donor formation dominates by a factor of 2 to 4. A more quantitative analysis of oxygen induced donors was not made due to difficulties in measuring the low levels of contamination involved. It is important to note at this point that the change from Hicks and Greene's quartz crucible to a graphite reactor introduced a shallow carbon acceptor. The transport of C by evolved oxygen in the form of CO into the epitaxy was not appreciated at that time. Deep Level Transient Spectroscopy (Lang (1974)) analysis would have revealed such a defect in a simple and routine way had it been available. Vilms and Garrett did note an acceptor impurity, the concentration of which varied with temperature, its identity was not established. Vilms and Garrett reported an activation energy of 1.52 eV, consistent with either vacancy formation or substitutional impurity incorporation. They speculated that the acceptor is either a native defect formed during the epitaxial growth, or that there is residual impurity with a large segregation coefficient in the solution and a consequent strong dependency on growth temperature.

Since Vilms and Garrett's main intention was to grow epitaxial layers in the 0.5 to 20 µm thickness range, particular care had to be placed onto uniform nucleation of the epilayer and onto the removal of the structure from the solution at termination of growth. Surfaces were cleaned, lapped and chemically-mechanically polished with a bromine-methanol etchant on an inert polishing pad. Finally, the samples were given a light etch in $H_2SO_4 : H_2O_2 : H_2O = 3:1:1$, currently a standard etchant for GaAs.

Graphite related growth arrangements were followed by the invention of the sliding-boat technique by Hayashi et al (1969) and Blum and Shih (1971). Here, the substrate is positioned in a machined graphite holder that can slide to contact in sequence several wells containing the saturated Ga solution. This method can give rise to a form of "volume-limited" growth, because the wells contain only a very small amount of the solution and spontaneous nucleation is thereby avoided.

Most LPE growth techniques involve some predetermined rate of cooling of a saturated gallium melt. In general these techniques are suitable for thin layers (~ 100 µm) but fail for thicknesses of more than a few hundred microns. The inherent reason is that the temperature interval used dictates the thickness limit. Disadvantages are also found in the doping profiles which vary as the growth proceeds as a consequence of temperature dependent segregation coefficients. For such reasons attention has been directed to the travelling solvent growth techniques first developed by Miavsky and Weinstein (1963). A modified version of this technique was introduced by Hesse et al (1972). Simply put, a temperature gradient about a constant mean temperature transports the dissociated As from the GaAs feed material to a substrate via a gallium solvent.

In this case the temperature profile is maintained by the axial temperature distribution of the furnace and the location of the crucible. The placement of the feed material in relation Co the substrate now clearly becomes a critical parameter. In general, LPE layers produced by such arrangements have shown poor crystallinity, suggesting that constitutional supercooling occurs due to insufficient temperature gradient. However, layer thickness exceeding 600 µm were achieved over a four hour period. Radiation detector diodes constructed in the form of a series of 'dots', were found to vary in both resolution and leakage currents suggesting to Hesse et al considerable lateral inhomogeniety in the epitaxial layer.

The nature of the previously mentioned anomalous layer also known as the i-layer has been of continuing interest, it is a high resistance epitaxy to substrate interface layer found in LPE and vapour phase epitaxy (VPE). This layer tends to be in the region of 0.2 to 2 μm wide and, because it includes a significant dip in the net carrier concentration profile, exhibits capacitive attenuation when fabricated into a device. This detrimental effect of the anomalous interface is well known in the construction of surface barrier radiation detectors and Gunn oscillators.

Blocker et al (1970) produced a detailed study of the interface layer by scanning electron microscopy and a series of capacitance-voltage (CV) measurements. They found that this region has a typical net acceptor density ranging from $10^{14}$ to $10^{16}$ cm$^{-3}$ which can be explained simply as an alteration in the balance of the net carrier concentration across a step transition from n+ to n$^-$. Since the substrate is nearly compensated so that $N_D$ and $N_A$ are both very much greater than /$N_D$ - $N_A$/ then an amphoteric impurity such as Si can alter this balance towards an acceptor state under a strain or temperature gradient. Similarly, "natural" impurities such as C could become electrically active by changing their lattice position. Alternatively, surface preparation of the substrate could leave a p-type impurity (e.g. Cu) at the liquid-solid interface growth front. Support for the last possibility came from photoluminescence studies of the interface layer by Nakashima and Hiras (1970) who found an emission band due to Cu acceptors at the interface layer. Further insight was gained by Di Lorenzo et al (1971) using direct image mass analysers and finding that these regions contain high concentrations of localised Si impurities with lesser concentrations of Li, Al and Fe. Further substance to this claim was given by Gibbons et al (1972) who suggested that poor device (diode) characteristics were due to an abrupt change in the free carrier concentration related to an impurity gradient between the epitaxial layer and the substrate. They considered that a desirable solution would be to introduce a buffer layer about 20 μm thick between the substrate and epilayer. However, growing such a configuration with a buffer concentration of $10^{15}$ cm$^{-3}$ proved difficult with only one growth producing good devices. An expected improvement in the I-V characteristic was not found. The anomalous layer was further investigated by Tavendale et al (1972). They found that when a radiation detector (essentially a diode under reverse bias) was exposed to infrared illumination a marked improvement was noted in the stability of pulse height response with bias and multi-peaking resulting from capacitive attenuation near full depletion of the diode. This effect can be explained simply as an increase of the conductivity of the anomalous layer as deep level acceptor impurities located there are field-deionised. The detector then tends towards real operation. The deionising of the anomalous layer acceptor does not exclude the possibility that other deep level defects are present in the bulk of the epitaxy. These might include deep donors which are deionised and similarly reduce pulse height variation. A further comment by Tavendale et al on reducing the effect of the anomalous layer is by using much deeper depletion layer. The capacitance radio aspect of the two layers would then give a lower charge attenuation. This would require high purity and LPE layers in excess of 200 μm thickness.

A different method from transient systems based on Nelson dip growth is the steady-state growth developed by Long, Ballantyne and Eastman (1974). This method allows lower growth temperatures and arbitrarily thick layers. Growth is achieved by establishing an equilibrium at the growing interface with a fixed temperature gradient between the substrate and a source crystal of undoped GaAs. The driving force for transporting As across the solution to the substrate is the temperature gradient. Transport is either by diffusion or by a combination of diffusion and convection which may account for the poor thickness uniformity commonly found. However, the advantage of thicker layers was outweighed by poorer purity in the 1 to 2.5 x $10^{15}$ $cm^{-3}$ range and surface variations dominated by edge effects. The vertical growth geometry, Figure 5a) and 5b) has the crucible and the seed, saturated Ga and the. GaAs source are located at appropriate points in the vertical temperature profile.

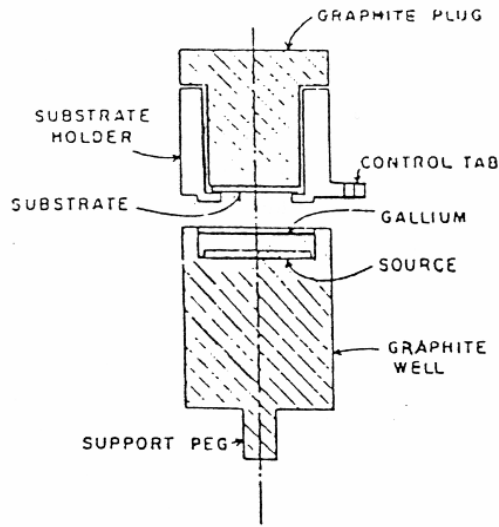

Figure 5a) –   Vertical steady-state boat in which the substrate is on top of the solution (after Long et al, 1974)

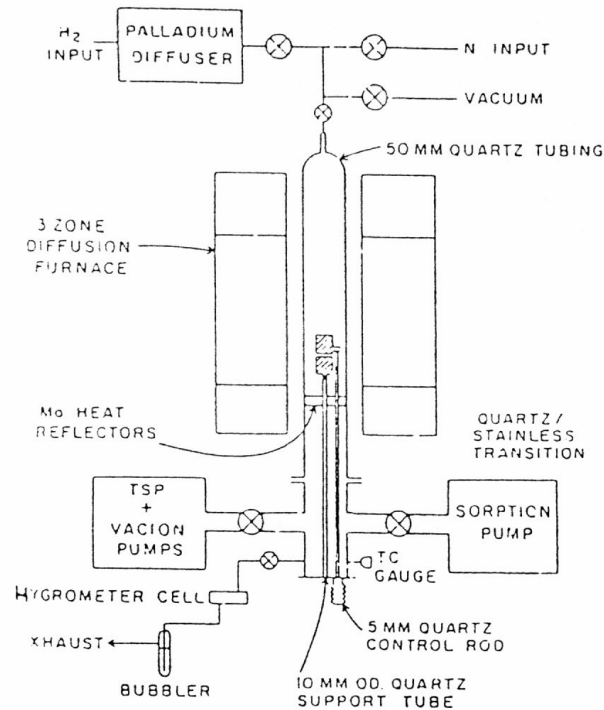

Figure 5b) – Vertical steady-state growth system (after Long et al, 1974)

Similar vertical growth arrangements were implemented by others, in particular Kobayshi et al (1976) used a simple dipping-type procedure. They eliminated poor control during the growth and especially over termination where Ga droplets often remain on the newly grown epilayer and result in uneven surfaces. The growth system consists of a graphite boat and the high quality palladium purified H . The resulting epilayers were of good quality with carrier concentration in the range of $10^{12}$ to $10^{14}$ cm$^{-3}$ . Charge trapping was noted and attributed to a deep acceptor produced by assumed Si contaminants, to reduce this adverse effect Fe was added (one part in three hundred) to the Ga melt. Excellent spectral results were obtained from surface barrier detectors built from such material.

Despite these numerous variations in LPE growth systems, defects and poor surface morphology resulting from uncontrolled microscopic growth velocity remained a problem. Joffe (1956) and Pfann et al (1957) suggested novel method to influence and control growth of an epitaxy in the immediate vicinity of the interface. The arrangement is a. typical vertical growth configuration modified to permit passage of an electric current. In such a way Pettier cooling (or heating) is introduced at the substrate-liquid interface. Kumagawa et al (1973) sucessfully used such a method, known as liquid-phase electroepitaxy (LPEE).

This earlier work was further developed by Jastrzebski et al (1978), (1980) and (1986) who produced a number of publications on growth kinetics in LPEE. Their method achieved bulk crystal growth up to 4mm thickness over a 20mm diameter wafer. The structure was of high quality, being essentially dislocation free. The net carrier concentration for such material could not be reduced below $10^{15}$ cm$^{-3}$ suggesting that passage of an electric current adversely affects impurity segregation in the growth front of the epitaxy. Further work on purification of LPEE layer by Bryskiewicz et al (1978) produced lower carrier concentrations in the order of $10^{14}$ cm$^{-3}$. Photoluminescence (PL) was used to identify the Si acceptor as the dominant residual impurity.

This was uniformly distributed through the LPEE, again confirming that segregation can be adversely affected by the electric current through the melt.

## RECENT ADVANCES IN GaAs MATERIAL

In this section, advances in GaAs material growth, characterisation and device fabrication are reviewed. These advances reflect the considerable R&D effort now being applied to GaAs, especially for optoelectronic, microwave and fast-logic devices.

To accommodate large volume production from single wafer GaAs, some significant improvements have been made in the growth of high resistivity "semi-insulating" ($\rho \sim 10^8$ $\Omega$–cm) undoped material using the liquid encapsulation Czochralski (LEC) pulling technique. Sumitomo Electric (Japan) have been able to produce commercially very low dislocation ($\sim$ 1000 cm$^{-2}$) material, up to 3 inch diameter, which is ideal for epitaxial substrates. Reduction of Si contamination is also achieved by using pyrolytic boron nitride (PBN) crucibles as reported by Shimada et al (1984).

Continued improvements in the purity of epitaxial GaAs grown by means other than LPE, such as molecular beam epitaxy (MBE) have been reported. MBE layers have been grown undoped at a carrier concentration of $2 \times 10^{14}$ cm$^{-3}$, by Hwang et al (1983) but unfortunately, MBE growth is intrinsically slow ($\sim$ 1 µm/h) and therefore applicable only to thin layer growth. Andrews (1983) and Abrokwah (1983) have used vapour phase epitaxy (VPE) and produced very low carrier concentrations ($10^{12}$ - $10^{13}$ cm$^{-3}$) but again such growth techniques only produce good quantity material at low growth rates of about 1 µm/h. Recent work by Boucher et al (1987) into newly developed LPE electroepitaxy (LPEE) have reported, by way of contrast, the possibility of growing quality "bulk epitaxial crystals" up to 4mm thick with low dislocation densities but relatively poor carrier concentrations, mostly around $1 \times 10^{15}$ cm$^{-3}$ with a best figure of $2 \times 10^{14}$ cm$^{-3}$.

GaAs has electrical properties which are strongly dependent on variations in stoichiometry which in turn depend on growth mode. For example, the dominant deep level defect, labelled EL2, is a deep donor having an activation energy of 0.82 eV to the conduction band. This defect performs a crucial role in the compensation of shallow acceptors due to carbon incorporated during the production of 'undoped' semi-insulating GaAs substrates. These are now used by the electronics industry as an alternative to Cr-doped semi insulating material.

Unfortunately, deep levels remove minority carriers either by trapping or by enhancement of recombination rates, so that the EL2 defect has a deleterious effect on the performance of GaAs devices, particularly nuclear radiation detectors. The EL2 defect has therefore been the subject of intensive investigation and considerable controversy. An understanding of its structure is central to the development of ultra high speed GaAs circuits.

The EL2 defect was originally thought to be oxygen-related (Lagowski et al (1984)) but electron paramagnetic resonance (EPR) studies have indicated that although the defect is present in bulk and VPE GaAs grown under As-rich conditions, it is absent in Ga-rich LPE, material. The defect can be detected by optical absorption or DLTS techniques, reported by von Bardeleben et al (1986) and by Alexiev and Tavendale (1984). Optical methods have been used by Holmes et al (1983) to map EL2 contours in LEC GaAs and show that EL2 formation is also enhanced by crystal stress (reflected by dislocation density). In a recent detailed investigation von Bardeleben et al (1986) attempted to identify EL2 using EPR and DLTS, proposing that the defect is the complex of the antisite defect As and an intrinsic interstitial Ga defect. As , with As in the first or second-nearest neighbour sites relative to As corresponding to the metastable and stable EL2 forms respectively. This description of the EL2 defect satisfied important observations made by Levinson (1983) of charge-state-controlled structural relaxation of the centre and intercentre optical transitions.

There has been considerable interest in the nature of irradiation-induced defects in GaAs, particularly as an aid in determining the structure of, for example, the EL2 defect (Pons and Bourgoin (1985), Stievenard and Bourgoin (1986). Further, the role of thermally induced defects, again involving the dominant EL2 deep donor defect, has recently been demonstrated by Lagowski et al (1986) in so-called inverted thermal conversion (ITC) material. Here, LEC GaAs (conducting or semi-insulating) is first subjected to a high temperature (1100-1200°C) anneal and fast cooling (quenching), which leads to a considerable reduction in the concentration of the EL2 defect, typically to less than $10^{15}$ cm$^{-3}$. A second anneal at 800°C (30 minutes) restores the EL2 defect and associated compensation giving high resistivity (2 x 10Ω – cm) n-type material. Kobayski et al (1976) had shown much earlier that it is possible to cycle reversibly between semiconducting and semi-insulating GaAs using LEG grown material having low C and Si concentrations by either slow-cooling or quenching from 950°C This effect also involves the EL2 defect compensation of residual acceptors. It is worth noting that these observations present the possibility of thermally controlling the conductivity of GaAs for radiation detector applications but it must also be kept in mind that the presence of a significant concentration of EL2 (or any other trap or records centre) will inevitably lead to poor detector performance, seen usually as asymmetric spectral lines with poor resolution. It appears that the most likely application for ITC - GaAs will be as substrate material.

Neutron transmutation doping of GaAs offers an alternate route for compensation of conducting material. The technique was first demonstrated for Si by Cleland et al

(1950) and is now well established in the silicon industry for production of uniformly phosphorus-doped material from float-zone single crystals via the reaction:

$$^{30}Si + n \longrightarrow {}^{31}Si \longrightarrow {}^{31}P + \beta$$

In the case of GaAs. due to the multiplicity of the natural isotopes of Ga and As the situation is more complex but essentially Ga and As transmute to Ge and Se donors which are electrically activated by thermal annealing. The radioactive ft decay period remains reasonably short.

| Transmutation Reaction | Capture Cross Section for Thermal Neutrons (barn) | Natural Half-Life | Abundance of GaAs |
|---|---|---|---|
| $Ga^{69} \longrightarrow Ga^{70} \longrightarrow Ge^{70}$ | 1.68 | 21 min | 60% |
| $Ga^{71} \longrightarrow Ga^{72} \longrightarrow Ge^{72}$ | 4.86 | 14 h | 40% |
| $As^{75} \longrightarrow As^{76} \longrightarrow Se^{76}$ | 4.30 | 26 h | 100% |

It is surprising that the technique was not applied to GaAs until 1970 (Marianashvii and Nanobashvili) to be followed later in a detailed report by Vesaghi (1982). Recently, studies of NTD treated semi-insulating Czochralski-grown GaAs have been reported by Mueller et al (1980) and Kolin et al (1984), NTD doped bulk-grown by Vigdorovich et al (1981) and Alexiev (1987). It appears that in thermal-neutron transmuted GaAs, radiation damage annealing commences at about 500°C and is completed at about 800°C (Mueller et al (1980) and Yahagi et al (1984)). The anti-site As defect is primarily involved in the annealing process (Schneider and Kaufmann (1982)). The fact that target doping can be attained at reasonably low temperatures for GaAs is encouraging, given the decomposition associated with high temperature treatment of the material.

As in the case of NTD-Si, it should be possible to at least increase the compensation in GaAs by a factor of 10, providing the inhomogeneity in the initial doping is no greater than about 10%. Thus, there is some prospect of reducing the lowest bulk-doped (p-type) GaAs presently available from $N \sim 10^{15}$ cm$^{-3}$ to $N \sim 10^{14}$ cm$^{-3}$ by the NTD technique. However, this doping level would still be too high for detector applications, and therefore low-doped LPE GaAs as the starting material, preferably with a $N \sim 10^{13}$ to $10^{14}$ cm$^{-3}$ range and p-type, becomes very attractive.

Passivation of both deep and shallow electrically-active defects by hydrogenation of GaAs has recently been the subject of intensive research. The discovery that a number of common deep levels in bulk, polycrystalline or LPE GaAs could be passivated following plasma-hydrogenation was first reported by Pearton (1982), Pearton and Tavendale (1982 and 1983). The EL2 level is also found to be deactivated on hydrogenation (Lagowski et al (1982)) with the electrical activity being restored by annealing at ~ 400°C. In lightly-doped n-type GaAs grown by MBE both shallow donors (e.g. Si) and the dominant deep level centres are neutralised by hydro-genation at 250°C with reactivation of the Si donors on annealing at 400°C and the deep levels at 600°C. Thus there is a temperature window within which it is possible to regenerate doping by shallow centres (controlling carrier concentration) yet still suppressing the deep level trapping centres (Dautremont-Smith et al (1986)). It is also interesting to note that the near-surface free hole concentration in p-type (Zn-doped) GaAs can be suppressed by hydrogenation, indicating neutralisation of acceptors (Johnson et al (1986)). Jalil et al (1987) and Pajot et al (1987) used infrared spectroscopy coupled with isotopic substitution (H substituted with D), and deduced that, in the cases of either, donor or acceptor neutralisation, the hydrogen is bonded to the As atom nearest to the dopant atom (Si or Zn) site.

The use of hydrogenation to passivate or neutralise remanent deep level defects in high purity LPE GaAs, while appearing to be a simple means of upgrading material quality, has the drawback that at most the process is only effective to a depth of a few pm, being regulated by diffusion rates. However, for applications such as passivation of surface-related defects, hydrogenation could well be useful in device fabrication.

In connection with the development of new GaAs device fabrication techniques, a considerable effort has been applied recently to the application of ion implantation doping, particularly for the construction of small dimensional channel or contact regions needed for high speed devices. This has in turn led to the application of new methods of dopant activation, e.g. rapid thermal annealing (RTA) and studies of the residual defects and solubility and activity of the implanted dopants. These topics have been extensively revised by Pearton et al (1987) and Williams and Pearton (1985). It is obvious that these techniques are also applicable to the fabrication of much larger devices such as formation of robust contacts on radiation detectors.

## PART 2

## *Liquid Phase Epitaxial growth of GaAs at the Australian Nuclear Science and Technology Organisation, Lucas Heights Research Laboratories.*

## *ABSTRACT*

Liquid phase epitaxial gallium arsenide layers, greater than 200 um thick and low net carrier concentration ( $N_{A,D} \sim 10^{13}$ cm$^{-3}$ ) have been grown in a silica growth system with silica crucibles. Analysis of electrical and chemical defects was carried out using deep level transient spectroscopy (DLTS) and secondary ion mass spectroscopy (SIMS), Details of the growth procedure are given and it is shown that silicon incorporation in the growth layer is not suppressed by the addition of ppm levels of oxygen to the main hydrogen flow; but appears to only suppress its electrical influence by residual, shallow acceptor -shallow donor net compensation.

## *Introduction*

Silica is a preferred high temperature containment material for the growth of a variety of semiconductor materials. For high purity liquid phase epitaxial (LPE) GaAs growth, silica is widely used for reactor tube construction. However, even though silica has a high free energy of formation it can not be regarded as inert. Silica when reduced ( or vitrified ) by reaction with GaAs melt will inject free Si into the growth front of a crystal and act as a dopant. Despite theoretical approaches such as that of Weiner [1], whose work centred on local thermodynamic equilibrium conditions, or the experimental approach by Hicks and Green [2], the problem of silicon contamination in open flow growth systems has been an ongoing concern; Si control has remained an uncertain factor in the growth mechanisms of LPE GaAs.

It is the purpose of the work described here to examine, for the case of high purity LPE GaAs growth, Si dopant interaction with the melt, the crucible and the ambient gas flow (Pd diffused $H_2$) using oxygen as a deliberately added impurity. The analysis of the epitaxial GaAs relies heavily on secondary ion mass spectroscopy (SIMS) and deep level transient spectroscopy (DLTS), techniques which were not available when earlier research interest was focussed on the growth of high purity liquid phase epitaxial materials. C-V profiling was used to determine the net carrier concentrations of the epitaxial layers.

The notation $SiO_2$-$SiO_2$-$H_2(O_2)$ denotes an open flow silica growth reaction tube ($SiO_2$-), with a silica crucible (-$SiO_2$-); $H_2(O_2)$ indicates a Pd diffused hydrogen gas stream with controlled addition of oxygen. For the work described here the silica crucible is loaded with a commercial GaAs substrate and a gallium melt saturated with GaAs. The containment of the liquid Ga melt in the silica crucible can be described by the following equilibrium reaction:

$$4Ga_{(e)} + SiO_{2(c)} \quad <=> \quad Si_{(s)} + 2Ga_2O_{(v)} \quad (1)$$
melt   crucible          dopant    vapour product

Cochran and Foster [3] have found that equation (1) is only an initial reaction between quartz and gallium producing Si. They indicated that $SiO_2(c)$ will continue to dissolve in Ga only until its concentration has reached a level where upon any further Si production results in a second reaction between the crucible and the silicon producing silicon monoxide

$$Si_{(in\ Ga)} + SiO_{2\ (c)} \quad <=> \quad 2\ SiO_{(v)} \quad (2)$$

Thus, a steady-state reaction between the gallium and silica can be described

$$2Ga_{(e)} + SiO_{2(c)} \quad <=> \quad SiO_{(v)} + Ga_2O_{(v)} \quad (3)$$

A companion reaction of the gallium liquid with the $H_2(O_2)$ gas flow may also occur:

$$2Ga_{(e)} + H_2O_{(v)} \quad <=> \quad Ga_2O_{(v)} \quad (4)$$

It should be noted that the product of equation (4) will affect both the equilibrium state of equation (1) and equation (3) by the production of $Ga_2O(v)$. However the dominant product capable of interacting with the melt is found in equation (1). There the free Si will be injected into the Ga melt and incorporated with the epitaxy. Si is well known to have a segregation coefficient of $k_{si} = 1.2$, and when residing on an As site -which is the dominant incorporation site for Si in Ga rich GaAs - will act as a shallow acceptor producing p-type GaAs. A second important impurity is oxygen, which is the dominant variable species, introduced deliberately into the $H_2$ stream and producing a shallow donor state in Ga rich GaAs (as opposed to a deep level in As rich GaAs) compensating the $Si_{As}$ shallow acceptor.

Thus, in summary, we note that the equilibrium state of equations (1) and (2) may be influenced by:
(a) the rate of removal of $Ga_2O_{(v)}$, so that the Si production rate may be altered by changing the gas stream velocity or the furnace temperature,

(b) controlled introduction of the oxygen species which can clamp $Ga_2O_{(v)}$ production by holding equation (1) near steady state equilibrium so that Si production remains low. Further introduction of oxygen may even shift equation (1) to the left hand side so that no further Si is produced,

(c)  using an appropriate growth (bake-out) temperature regime: the distribution coefficient of oxygen and ultimately the donor concentration has been shown by Otsubo et al [4] to decrease linearly with increased bake-out temperatures.

(d)  the production and retention of electrically inactive $SiO_x$ species in the melt influenced by the rate removal of $SiO_{(v)}$ and the concentration of $Si_{(s)}$.

The importance of the above interactions for the growth of high purity GaAs is examined in the following sections.

## *The epitaxial growth system*

The growth system consists of a typical horizontal tilt-tube furnace of the type first described by Nelson [5]. Inside the tube a flow of hydrogen is maintained at near ambient pressure. To produce high purity liquid phase epitaxial gallium arsenide particular emphasis has to be placed on the quality of the hydrogen atmosphere used in the open-flow growth system. As with all such growth systems, unpurified $H_2$ is passed through a palladium diffuser (Resource Systems Inc DSPS-1) removing numerous gaseous impurities such as gaseous carbon ($CO_x$), hydrogen sulphide, varying levels of water vapour and oxygen, all of which could constitute electrically active dopants in gallium arsenide.

The quartzware is of high purity "Spectrosil" silica, the boat (crucible) is also of "Spectrosil" grade silica. Other crucible materials have been examined elsewhere [6]. The hydrogen flow can be controlled between 0 and 2 standard litres per minute (SLM). The water content of the hydrogen is controlled by bleeding minute (ppm) quantities of oxygen between the $H_2$ purifier and the furnace using a Granville-Phillips leak valve type 203. The leak valve has a resolution of 1/10 ppm, but setting reproducibility is poor and once a working level is found further adjustment to the $O_2$ level is achieved by varying the $H_2$ flow rate, thereby altering the dilution ratio of the two gases. Most experimental levels of $O_2$ in $H_2$ would be within the range of 0 to 10 ppm of which 1.6 to 2.2 ppm in steps of 0.1 ppm of $O_2$ would be typical. When no $O_2$ input is required then a fixed 'cajon' seal is used to block-off the supply. The amount of oxygen passing through the leak is measured at the gas output of the furnace using a selfcalibrating SYSTECH oxygen analyser model 2550. The $O_2$ measuring response is of the order of a few seconds and can be calibrated using the $O_2:N_2$ ratio of air. A recorder output provides continuous $O_2$ level readings over an experimental period which may last over four days from start to finish, of which 48 hours is required to obtain a steady state condition for the $O_2$ (ppm)/$H_2$ ratio.

## *Furnace design and control*

The furnace design for epitaxial growth has to meet some specific requirements. It must have minimal or no temperature gradient over at least the length of the growth crucible, so that the melt and seed (substrate) are at the same temperature. A radial temperature gradient must be maintained so that when the melt is over the seed, the latter remains cooler. If this is not the case, partial or total melt-back of the seed may occur with indiscriminate nucleation of the regrowth. Finally assembling and servicing the growth system must be easy: the furnace must be detachable to expose the silica reactor tube which has to be removed routinely for etching.

For these reasons the furnace was designed so that it consists of essentially two hemicylindrical shapes, of 425 mm length and 210 mm diameter. The outer skin is of copper over which a 6 mm diameter copper tube is soldered in a serpentine fashion providing heat removal with circulating tap water. The furnace cavity is insulated with silica wool.

The heating element is located in the top section of the furnace and is similarly hemicylindrical in shape being 305 mm long and 63 mm in diameter wound as a single element spirally along the inner face of the element former. The element wire used was 10G Kanthal "A" and the cold resistance of the element was 12.5 Ω. This design proved to be long-lasting with an average life of 10 months.

The longitudinal temperature gradient is only 1 - 2° $C.cm^{-1}$ within the region where the crucible is resident. While the vertical radial temperature distribution is 14° $C. cm^{-1}$.

Power control to the furnace is achieved with a Leeds and Northrup Electromax III controller and a Leeds and Northrup type 11903 zero voltage power package. All thermocouples used were Pt-Pt 13 % Rd located in a twin bore alumina tube.

## *Substrate preparation*

Substrates used as 'seeds' for all epitaxial growth experiments were obtained from M. C. P. Electronic Materials Ltd UK. All substrates were oriented to the (100) crystallographic plane, had a net free carrier concentration between 0.7 and $1x10^{16}$ $cm^{-3}$ and were horizontal Bridgman (HB) grown. One face was mechanically polished by the supplier. Immediately before each growth, the substrates were degreased in xylene with ultrasonic agitation for 10 minutes followed by displacive rinsing in methanol and further rinsing in 18 MΩcm $H_2O$. After the degreasing-cleansing procedure the substrates were etched in $3H_2SO_4:H_2O_2:H_2O$ at 100° C for 2 minutes to remove microscopic surface damage produced by the polishing operation. Displacive rinsing in 18 MΩcm $H_2O$ followed the etch after which the substrate was given a further etch in $HCl:H_2O$ for 10 minutes to remove the soft amorphous oxide layer left on the GaAs surface. Then finally, after a last displacive rinse in 18 MΩcm $H_2O$, the substrate was blown dry with filtered nitrogen gas and immediately loaded into the prepared crucible.

## *Saturation of the Ga:GaAs melt*

The Ga:GaAs melts prepared for epitaxial crystal growth were made from nominal 7N (99.99999) pure Ga supplied by Alcan Corp. in 25g ampoules. The bulk GaAs used for saturating the Ga was supplied by MCP Electronic Materials with a net free carrier concentration of ~ 1 x $10^{16}$ cm$^{-3}$. To prepare a melt for growth, the Ga:GaAs solution must be saturated at the starting growth temperature otherwise total loss or deep melt-back of the seed will occur. However, at the bake-out temperature the solution has to be undersaturated so that all the feed material, the bulk GaAs, is in-solution. SIMS analysis of the bulk GaAs indicates a Si content of 2 x $10^{16}$ cm$^{-3}$ (figure 1). Once this is in solution it can be controlled as described later. It will also be appreciated that a higher growth starting temperature requires larger quantities of GaAs for saturation and consequently thicker epitaxies can be grown. A schematic Ga-As phase diagram indicating an increase of As solubility in Ga (X) from $X_C$ to $X_A$ as temperature increases from TB to TA (9 to 14 at at 850 to 960° C) is given by Dawson [7].

To determine the bulk GaAs weight required to saturate the Ga melt (usually 50g) at a particular temperature, two methods can be used [7], one numerical and the second, the more convenient graphical method. The graphical method is however derived from the numerical method so that only the latter is descibed here:

Method:  Given X'$_{As}$ (the value X'$_{As}$, at a particular saturation temperature can be found from standard solubility curves [7])
 and gallium mass, m$_{Ga}$, (usually 50g) we require to find the mass of added GaAs for saturation,
Thus, the value m$_{GaAs}$, can be found from the following:

$$X'_{As} = \frac{n_{As}}{n_{total}}$$

$$= \frac{n_{GaAs}}{n_{Ga} + 2n_{GaAs}}$$

$$= \frac{\frac{m_{GaAs}}{M_{GaAs}}}{\left[\frac{m_{Ga}}{M_{Ga}} N_0 + 2 \frac{m_{GaAs}}{M_{GaAs}} N_0\right]}$$

$$\therefore m_{GaAs} = 2.07 \frac{X'_{As}}{1 - 2X'_{As}} m_{Ga}$$

where $m = \dfrac{m_{GaAs}}{M_{GaAs}} N_0$

($N_o$ = Avogadro's number and $X'_{As}$ = mole fraction of As in Ga at a particular temperature).

For example:
$M_{Ga} = 69.7$ g/mol
$M_{AS} = 74.9$ g/mol
$M_{GaAs} = 144.6$ g/mol

Thus, for a melt of 50g Ga to be saturated at 830°C 5g of GaAs will be required, though often an extra 2 g is usually added to ensure that the melt is not undersaturated due to small variations in the tip temperature. Note that it is highly desireable to supply the Ga from small ampoules since decanting of a large volume of Ga can lead to contamination through heating and reheating of the solidified Ga.

## Silica crucible preparation

The Silica used for this type of crucible is "Spectrosil" a synthetic silica supplied by Thermal Syndicate, UK. Total metallic impurities are less than 0.02 ppm and at least a factor of 100 better than the more common "Vitreosil" silica available. "Vitreosil" can contain as much as 2.5 ppm Fe and 4 ppm Li, both of which are electrically active impurities in GaAs.

Preparation for epitaxial growth involves first welding a silica whisker, also of "Spectrosil" grade, to hold down the substrate which would otherwise float on the gallium melt. This is followed by a normal silica cleaning procedure: degreasing in xylene, rinse in methanol followed by washing in 18 MΩcm $H_2O$. Etched for 10 minutes in $HNO_3$:HF =4:1 at room temperature, rinsed and re-etched in HC1. HC1 is used to remove Au replating from HF known to be a common contaminant. Finally, the etched crucible is rinsed in 18 MΩcm $H_2O$ and dried with N2 gas. This preparation usually coincides with immediate loading of the gallium melt and substrate.

## Experimental Technique

Growth is commenced by tilting the furnace and thereby flowing the saturated gallium melt over the substrate. Preparation of the crucible and the saturation of the melt is described above with great emphasis placed on minimising contamination during etching, substrate preparation and loading of the crucible.

The furnace cooling rate was programmed simply with a multiratioed gear box and a motor driven externally with attached helipots which alter the set point conditions on a temperature controller. Temperature run-down rates explored were 0.3 to $100^0 C.H^{-1}$ with a preferred rate of $7°C.H^{-1}$ resulting in a growth rate of 43 $\mu m.H^{-1}$, spanning over 7 hours and producing good crystallinity. Vibrational stirring was used during growth as a means of improving GaAs homogeneity in the melt, which subsequently improves the crystallinity of the LPE layers [8].

The Nelson tilt type furnace proved to be easy to use and adaptable to a variety of varying experimental conditions; the overriding necessity was that the loading of the melt and substrate into the crucible must be simple. Similarly loading of the crucible into the furnace also had to be straightforward. Complications in handling procedure can lead to contamination of the melt and crucible; long periods of substrate exposure to air will oxidise its surface resulting in uneven substrate melt-back, poor crystallinity and unsatisfactory surface morphology [8].

There are two distinct phases of temperature control during crystal growth, first there is a preheat of the melt, referred to as the bake-out period - this can occur at a temperature independent of the growth temperature. Then there is the actual growth phase which occurs over a temperature range beginning at the temperature at which the melt is tipped over the substrate (referred to as the 'tip' temperature) at the start of crystal growth, and ending at a lower temperature at the conclusion of growth where upon the GaAs:Ga melt is removed from the substrate.

The bake out temperature used was 850° C for 14 hours followed by a tip temperature of 830° C and a temperature run down of 7° C per hour. Melt saturation was always calculated so that the feed GaAs completely dissolved at the bakeout temperature. Thus the melt would be just under saturated at bakeout, whilst at the growth tip temperature the melt would be saturated with some recrystallisation on the surface of the liquid gallium melt.

Particular attention was given to the stability of the $0_2$ level before and on completion of an epitaxial growth. In this way 0.1 ppm of $0_2$ resolution could be expected.

## Results and Discussion

All the epitaxies grown for this study were examined using DLTS with no deep level traps being detected - this is a well known consequence of LPE growth in the $Si0_2$-$Si0_2$-$H_2(0_2)$ system [9]. These results indicate that the silicon and oxygen impurities are incorporated as shallow levels. The LPE material grown by us has also demonstrated extremely high electrical purity as found from minority carrier diffusion lengths measurements [10]|, and as evidenced by the successful use of this material by Alexiev and Butcher [11] for the construction of room temperature Schottky barrier nuclear radiation detectors of x-rays and low energy y-rays.

It was found from SIMS analysis that silicon was introduced almost uniformly from one experiment to the next regardless of the oxygen levels used in our system (figure 2a). It was also evident from SIMS analysis that the shallow controlling donor due to oxygen could be introduced to compensate epitaxies to a net carrier concentration as low as $2x10^{13}$ cm$^{-3}$ (figure 2b). However, C-V profiling of angle lapped epitaxial layers over 350 µm revealed that the compensation is not uniform throughout the thickness, but has a tendency to increase in n-type or decrease in p-type by a factor of approximately 3, as shown in figure 3a and 3b. The nature of the slope in N(x) versus W profiles can be explained firstly by the changing solubility of oxygen with temperature in the gallium melt. Ostubo [4] found that oxygen is more soluble in Ga at a lower growth temperature than at a higher growth temperature (ko = $5x10^{-5}$ at 800 to 725°C and ko == $6.5x10^{-4}$ at 700 to 625°C). Added to this is a decrease in As solubility in Ga with temperature decrease. Both these effects will increase the formation of the shallow donor.

The results gained from SIMS analysis ( figure 2a ) indicate that Si is introduced during the bake-out period virtually at a steady level between $10^{15}$ to $10^{16}$ atoms cm$^{-3}$. Some small decrease in Si incorporation may be present near the point of compensation, however at the higher ppm levels of oxygen introduction no suppression of Si injection from the silica occurred. This contradicts the assumption of earlier workers who on the basis of electrical measurements alone had postulated such Si suppression. The SIMS results presented here demonstrate that this is clearly not the case. Further work is required to understand why Si suppression does not occur, as should follow from equation 1. For such work the role of the $Ga_2O$ and SiO vapour products needs to be clearly established by residual gas analysis of the hydrogen flow over the melt. The role of reactants within the melt which may be in equilibrium with these vapour products also needs to be clarified. For instance incorporation in the LPE epitaxy of neutral defects, such as the SiOx species, may be studied by activation through thermal annealing as has been demonstrated by Alexiev et al. [12].

## *Conclusions*

The following observations can be made when growing LPE GaAs in the $SiO_2$-$SiO_2$-$H_2(O_2)$ system: Si is incorporated in the gallium melt in contact with a silica crucible to an average $5x10^{15}$ atoms per cm$^{-3}$, supporting the involvement of the intermediate reaction step described by equation 1. Oxygen does not suppress Si incorporation from a silica crucible; but appears to only suppress its electrical influence by residual, shallow acceptor -shallow donor net compensation. $Si_{As}$ and oxygen ($O_{As}$) even in large quantities (ppm levels) in the epitaxy do not form detectable deep level traps in this gallium rich material. Finally high purity, low net carrier concentration epitaxies can be produced by precise control of oxygen.

A following report will describe minority carrier diffusion lengths for high purity liquid phase epitaxial GaAs and the fabrication of high purity liquid phase epitaxial GaAs nuclear detectors.

*Acknowledgments*

We would like to acknowledge our debt to the late Dr A.J.Tavendale who not only gave us the opportunity to commence this project, but also gave considerable encouragement and constructive suggestions during the earlier phases of this work. Our thanks to Prof. L.R.Dawson of Sandia National Labs Albuquerque New Mexico who introduced us to the topic of liquid phase epitaxial growth through the provision of an excellent set of lecture notes. Also our thanks to Prof S.J.Pearton of the University of Florida who was instrumental in arranging our SIMS measurements.

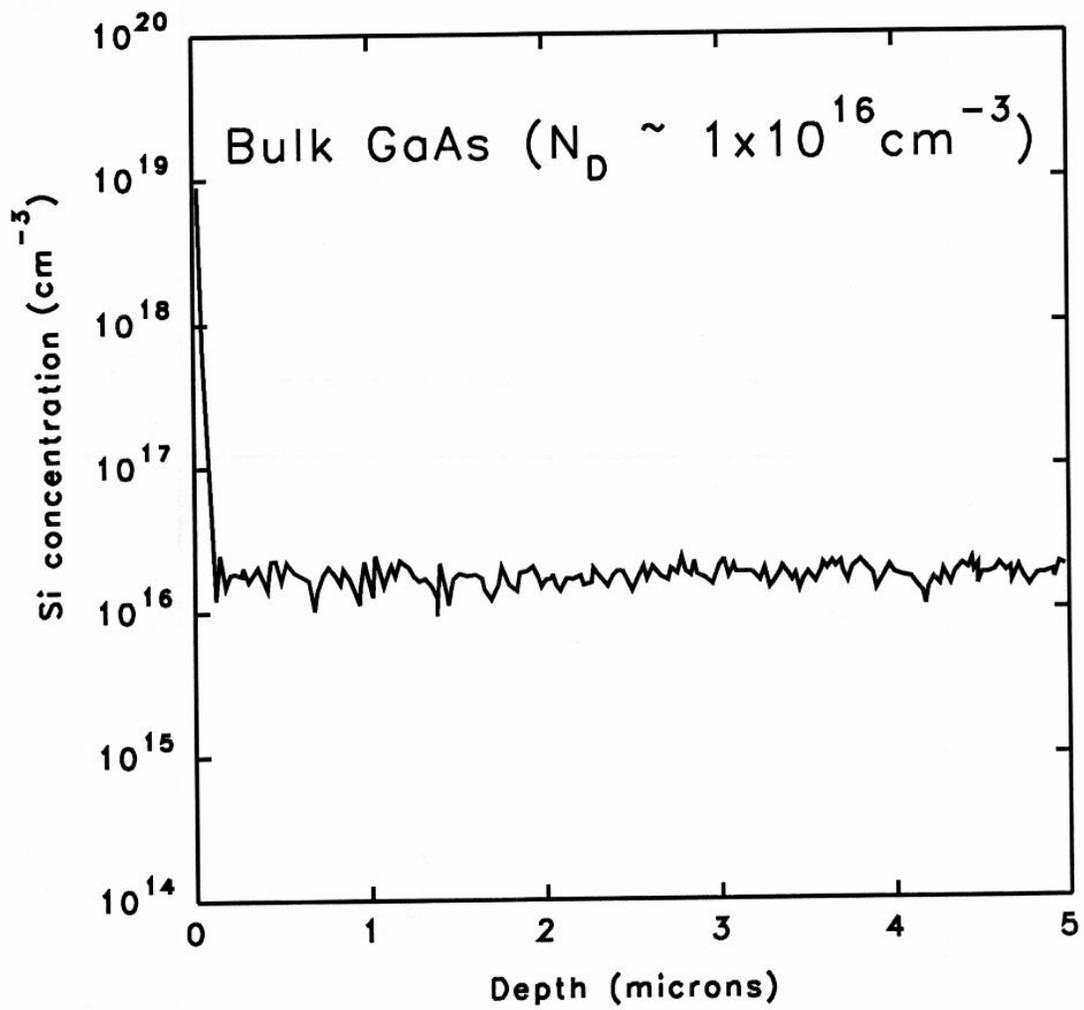

Figure 1: SIMS measurment of LPE GaAs indicating Si content in the epitaxy.

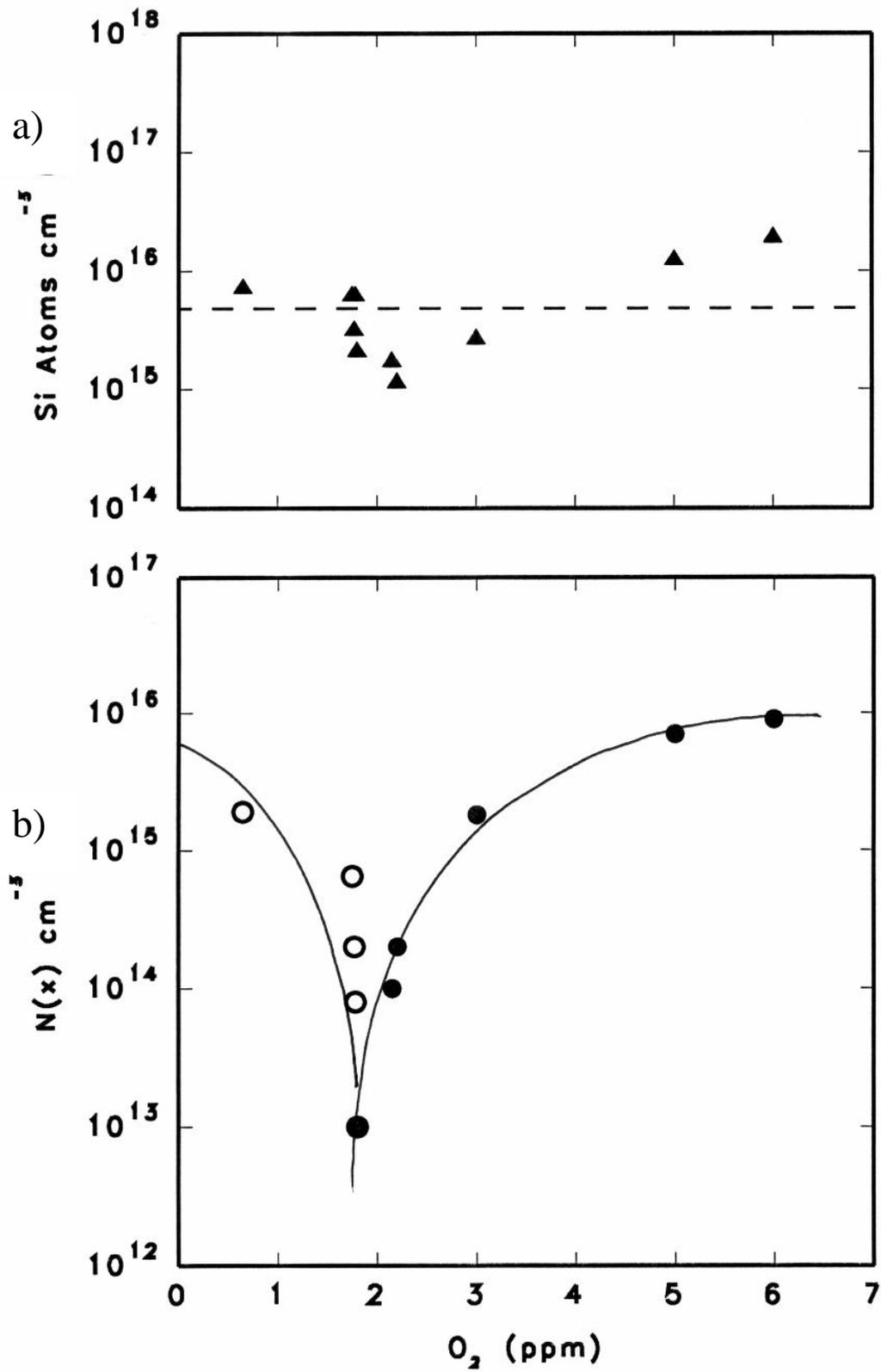

Figure 2: a) Si content in GaAs found by SIMS for various oxygen levels, b) N(x) values obtained for various oxygen levels.

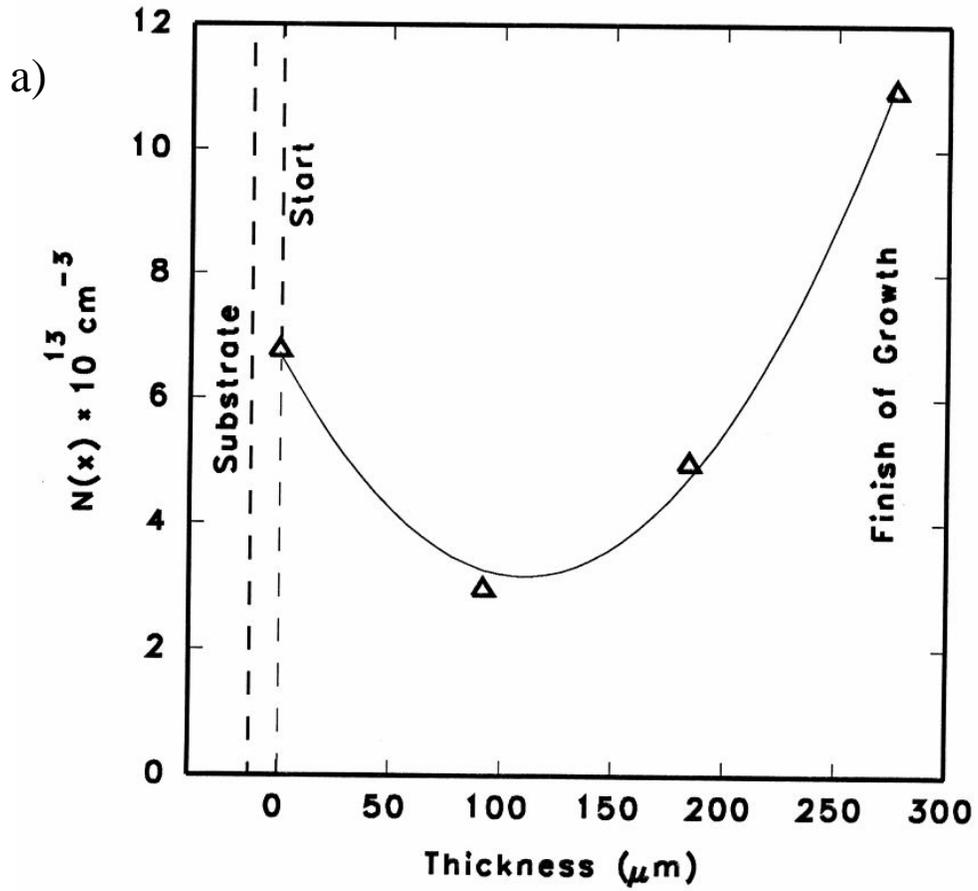

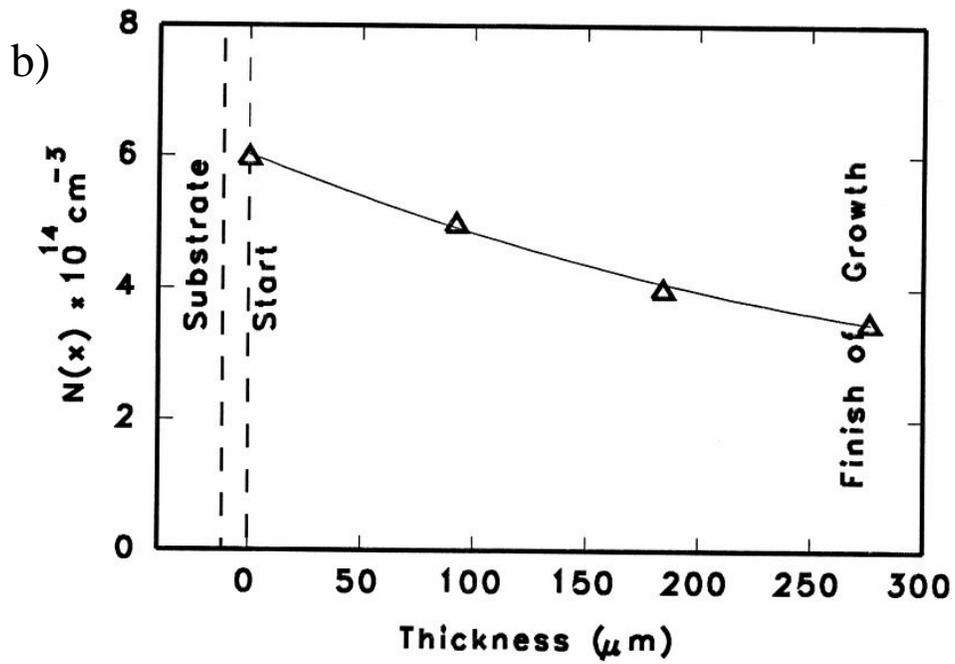

Figure 3: N(x) depth profiles for n-type (a) and p-type (b) epitaxies, indicating carrier concentration nonuniformity within the epitaxies.